\documentclass[12pt]{article}
\usepackage{epsfig}
\def\wtG{\widetilde G}
\def\ovl{\overline }

\def\ra{\rangle}

\begin{document}

\begin{flushright}
hep-ph/0301224\\
\end{flushright}

\begin{center}
{\Large\bf Isospin violation in
mixing and decays \\
of $\rho$- and $\omega$-mesons\footnote{
Talk presented at KTM-80, dedicated to 80th anniversary of
K.A.Ter-Martirosyan, Moscow, ITEP, 30~Sep.--1~Oct.~2002.}}

\vspace{0.4cm}

{\large Ya.I.Azimov }

\vspace{0.3cm}

{\it Petersburg Nuclear Physics Institute,\\
Gatchina, St.Petersburg, 188300, Russia}

\vspace{0.3cm}
\end{center}

\begin{abstract}
Influence of the isospin-violating $(\rho^0,\,\omega)$-mixing
is discussed for any pair of decays of $\rho^0,\,\omega$
into the same final state. It is demonstrated, in analogy
to the $CP$-violation in neutral kaon decays, that isospin
violation can manifest itself in various forms: direct
violation in amplitudes and/or violation due to mixing.
In addition to the known decays $(\rho^0,\,\omega)\to\pi^+\pi^-$
and $(\rho^0,\,\omega)\to\pi^0\gamma$, the pair of decays to
$e^+e^-$ and the whole set of radiative decays with participation
of $\rho^0,\,\omega$ (in initial or final states) are shown to be
also useful and perspective for studies. Existing data on these
decays agree with the universal character of the mixing parameter
and indirectly support enhancement of $\rho^0\to\pi^0\gamma$ in respect 
to $\rho^{\pm}\to\pi^{\pm}\gamma$. Future precise measurements will
allow to separate different forms of isospin violation and
elucidate their mechanisms.

\vspace{0.3cm}

\end{abstract}

\newpage

\section {Introduction}

It is widely known that the isospin symmetry {\it is} violated.
But nobody knows {\it why} and {\it how} it is violated. There
are at least two possible sources of the violation:
\begin{itemize}
\item QED does not respect the isospin, since different members
of any isomultiplet always have different electric charges. As a
result, the photon can be considered as a two-component object
with isospins $I=0,1\,$. Therefore, presence of photons, real or
virtual, inevitably spoils the symmetry. The corresponding effect
for processes without real photons is expected to be
${\cal O}(\alpha)$ in the amplitude.
\item QCD can also violate isospin, due to different properties
of $u$- and $d$-quarks. Most popular here are references
to different quark masses, but other properties, not always
directly related to masses, may also be efficient (as examples,
I can mention magnetic moments, or difference of quark wave
functions inside hadrons). Estimates of the expected effect
in such approaches are rather ambiguous.
\end{itemize}
Experiments demonstrate isospin violation ({\it e.g.}, hadron
mass differences) mostly at the relative level of order $10^{-2}$
or less. This does not allow even to discriminate between the two
above mechanisms. Thus, further studies, both theoretical and
experimental, are necessary to elucidate the underlying physics.

A favourable site for such studies may be provided by mixing of
$\rho^0$- and $\omega$-mesons, where some enhancement becomes
possible due to $M_\omega\approx M_\rho\,$. A well- and long-known
example is the decay $\omega\to\pi^+\pi^-$. The isospin symmetry
totally forbids it (initial $I=0$, final $I=1$), but the mixing
opens the cascade transition $\omega\to\rho^0\to\pi^+\pi^-$. The
resulting branching ratio achieves 2\%~\cite{PDT2}, instead of
${\cal O}(\alpha^2)$.

A more recent example of possible manifestation of the mixing is
given by decays $\rho\to\pi\gamma$. There are experimental
evidences for enhancement of the neutral decay in respect to
charged one (see \cite{PDT2}; exact value is still uncertain,
as evident from comparison of the corresponding numbers in
the neighbouring issues of Particle Data Tables~\cite{PDT2,PDT0}).
Meanwhile, the isospin conservation admits only the isoscalar
photon component to participate in those decays, and so probabilities
for $\rho^0\to\pi^0\gamma$ and $\rho^{\pm}\to\pi^{\pm}\gamma$ were
expected to be the same. Their inequality (either enhancement or
suppression of the neutral decay) may emerge from contribution of
the cascade $\rho^0\to\omega\to\pi^0\gamma$ which is impossible
for the charged decay (see~\cite{BOC} and references therein).

In a recent paper~\cite{EPJA}  I suggested to broaden the set of
decays under consideration, since any pair of the decays
$\omega,\rho^0\to$(the same final state) should be sensitive to the
$(\rho,\omega)$-mixing. This talk gives a brief presentation of ideas
and results of the paper~\cite{EPJA}.

\section {Vector meson mixing}

Let us begin with bare states $\omega^{(0)}$ and $\rho^{(0)}$.
They have bare (complex) masses
$$M_\omega^{(0)}=m_\omega^{(0)}-\frac{i}2 \Gamma_\omega^{(0)}\,,
~~~M_\rho^{(0)}=m_\rho^{(0)}-\frac{i}2 \Gamma_\rho^{(0)}$$
and bare propagators
\begin{equation}
[D_\omega^{(0)}(k^2)]_{\mu \nu}=
\frac{g_{\mu \nu}-\frac{k_\mu k_\nu}{M^{(0)2}_\omega}}
{k^2-M^{(0)2}_\omega}\,\,, ~~~~~
[D_\rho^{(0)}(k^2)]_{\mu \nu}=
\frac{g_{\mu \nu}-\frac{k_\mu k_\nu}{M^{(0)2}_\rho}}
{k^2-M^{(0)2}_\rho}\,\,.
\end{equation}
Mixing arises if there exist transitions $\omega^{(0)}\to\rho^{(0)}$
and $\rho^{(0)}\to\omega^{(0)}$. Corresponding transition vertices
may be described by transition amplitudes $G_{\omega\rho}$ and
$G_{\rho\omega}$ respectively\footnote{See~\cite{EPJA} for more
detailed description of the vertices.}. Summation over all mutual
transitions provides four different propagators for bare states:
$$D_{\rho\rho}(k^2),~~D_{\rho\omega}(k^2),~~D_{\omega\rho}(k^2),
~~D_{\omega\omega}(k^2),$$
which describe all reciprocal transformations of $\rho^{(0)}$ and
$\omega^{(0)}$. Together they may be considered as a $2\times2$
matrix propagator. Its diagonalization  picks out physical
propagators $D_\omega(k^2)$ and $D_\rho(k^2)$ with physical
masses
\begin{equation}
M_\omega^2= M^2+K\delta M^2\,,~~~~ M_\rho^2=
M^2-K\delta M^2\,, \end{equation}
where
$$\delta M^2=\frac{M^{(0)2}_\omega-M^{(0) 2}_\rho}2\,,
~~~M^2=\frac{M^{(0)2}_\omega+M^{(0)2}_\rho}2\,,$$
$$K=\sqrt{1+\wtG_{\rho\omega}\wtG_{\omega\rho}}\,,~~~~
\wtG_{\rho\omega}=\frac{G_{\rho\omega}}{\delta M^2}\,,~~~
\wtG_{\omega\rho}=\frac{G_{\omega\rho}}{\delta M^2}\,.$$

Now we can consider a process $i\to f$ where $\rho^0$ and/or
$\omega$ appear as the intermediate states. Its amplitude
in terms of bare states is
\begin{equation}
A_{if}=A^{(0)}_{i\rho} D_{\rho\rho}A^{(0)}_{\rho f}\,+\,
A^{(0)}_{i\rho} D_{\rho\omega}A^{(0)}_{\omega f}\,+\,
A^{(0)}_{i\omega} D_{\omega\omega}A^{(0)}_{\omega f}\,+\,
A^{(0)}_{i\omega} D_{\omega\rho}A^{(0)}_{\rho f}\,,
\end{equation}
where $A^{(0)}_{i\rho}, A^{(0)}_{i\omega}$ are production
amplitudes for bare $\rho^{(0)}$-, $\omega^{(0)}$-states, while
$A^{(0)}_{\rho f}, A^{(0)}_{\omega f}$ are their decay amplitudes.
The whole amplitude may be rewritten in terms of physical states
in the simple form
\begin{equation}
A_{if}=A_{i\rho} D_{\rho}A_{\rho f}\,+\,
A_{i\omega} D_{\omega}A_{\omega f}\,,
\end{equation}
where the physical propagators $D_\rho(k^2),~D_\omega(k^2)$
are used together with the physical amplitudes
\begin{equation}
A_{i\rho}=\sqrt{\frac{K+1}{2K}}\left(A^{(0)}_{i\rho}-
A^{(0)}_{i\omega}\frac{\wtG_{\omega\rho}}
{K+1}\right)\,,~~~
A_{i\omega}=\sqrt{\frac{K+1}{2K}}\left(A^{(0)}_{i\omega}+
A^{(0)}_{i\rho}\frac{\wtG_{\rho\omega}}
{K+1}\right)
\end{equation}
for the $\rho^0$-, $\omega$-meson production and
\begin{equation}
A_{\rho f}=\sqrt{\frac{K+1}{2K}}\left(A^{(0)}_{\rho f}-
\frac{\wtG_{\rho\omega}}{K+1}
A^{(0)}_{\omega f}\right)\,,~~~
A_{\omega f}=\sqrt{\frac{K+1}{2K}}\left(A^{(0)}_{\omega f}+
\frac{\wtG_{\omega\rho}}{K+1}
A^{(0)}_{\rho f}\right)
\end{equation}
for the meson decays.

The picture of mixed $\rho^{(0)}$-, $\omega^{(0)}$-states is
similar to the well-known picture of mixing for $K^0, \ovl K{}^0$,
as described by Lee, Oehme, Yang~\cite{LOY}. It corresponds to
diagonalization of the mass squared matrix of the
$(\rho,\omega)$-system
\begin{equation}
{\cal M}^2=
\left(\begin{array}{cc}
M_\rho^{(0)\,2}& G_{\omega\rho}\\
G_{\rho\omega}&M_\omega^{(0)\,2}
\end{array}\right)
\end{equation}
(and its matrix propagator ${\cal D}= (k^2-{\cal M}^2)^{-1}$ )
in the form
\begin{equation}
{\cal M}^2=
\sqrt{ \frac{K+1}{2K}}
\left(\begin{array}{cc}
1&\frac{\wtG_{\omega\rho}}{K+1}
\\
-\frac{\wtG_{\rho\omega}}{K+1}
&1\end{array}\right)  \cdot
\left(\begin{array}{cc}
M_\rho^{2}&0\\0&
M_\omega^{2}\end{array}\right) \cdot
\sqrt{ \frac{K+1}{2K}}
\left(\begin{array}{cc}
1&
-\frac{\wtG_{\omega\rho}}{K+1}
\\
\frac{\wtG_{\rho\omega}}{K+1}
&1\end{array}\right)\,.
\end{equation}
The bare states $|\rho^{(0)}\ra$
and $|\omega^{(0)}\ra$ appear to be analogs of flavour states
$|K{}^0\ra$ and $|\ovl K{}^0\ra$, while the physical states
\begin{equation} |\rho\ra=N_\rho\left(|\rho^{(0)}\ra-\,
\frac{\wtG_{\rho\omega}}{K+1}
\,|\omega^{(0)}\ra\right)\,,~~~
|\omega\ra=N_\omega
\left(\frac{\wtG_{\omega\rho}}{K+1}
\,|\rho^{(0)}\ra+\,|\omega^{(0)}\ra\right)
\end{equation}
play the role of $|K_S\ra$ and $|K_L\ra\,$ (compare with
expressions (6); $N_\rho$ and $N_\omega$ are normalizing factors).
The essential difference, however, is the nonvanishing $\delta M^2$,
which would imply $CPT$-violation in the case of $(K^0\ovl K{}^0)$.
As for the neutral kaons, there is a possibility of rephasing for
$\rho^{(0)}$ and $\omega^{(0)}$. $T$-invariance makes possible to
fix their phases so that
$$\wtG_{\rho\omega}=\wtG_{\omega\rho}\equiv \wtG\,.$$
Analogy between the two systems would be more evident if one
could observe oscillating time distributions of $\rho$- and
$\omega$-decays. This is, however, quite unrealistic, and we can study
only time-integrated double-pole distributions in $k^2$. More detailed
discussion of similarity and difference between ($\rho^0, \omega$) and
($K^0, \ovl K{}^0$) may be found in~\cite{EPJA}. I would like,
nevertheless, to mention here one unfamiliar point: while the bare
states are orthogonal, the physical ($\rho^0, \omega$)-states are
orthogonal only if $\wtG$ is real.

\section {Mixing and isospin violation in decays }

Symmetry violations in decays of neutral kaons are known to reveal
themselves in two forms: mixing violation, manifested in mixing
parameters of eigenstates; and direct violation, seen as a property
of one or another particular amplitude for kaon decays. Isospin
violation for the ($\rho, \omega$)-system may also have two forms.
It can be direct violation, seen in production or decay amplitudes
for bare states; or it can be mixing violation due to dimensionless
parameters $\wtG_{\rho\omega}$ and $\wtG_{\omega\rho}$.
Existing experience allows to expect relative effects in
amplitudes $\sim0.01$ for the direct violation, while $|\wtG|$ might
be up to 0.1.  This apparent enhancement of $\wtG$ arises due to the
denominator $\delta M^2$, small at the hadron mass scale. Nevertheless,
the difference is not very strong, and future accurate description may
require to account for the both kinds of isospin violation.

Let us compare a pair of decays $(\omega,\rho^0)\to f$ with the same
final state. The ratio of their amplitudes is
\begin{equation}
a_{\omega/\rho0 f}\equiv \frac{A_{\omega f}}{A_{\rho f}}=
a^{(0)}_{\omega/\rho0 f}\,
\left(1+\frac{\wtG}
{(K+1)\,a^{(0)}_{\omega/\rho0 f}}\right)\,
\left(1-\frac{\wtG\,a^{(0)}_{\omega/\rho0 f}}
{K+1}\right)^{-1}\,,
\end{equation}
where we assume $T$-invariance and define
$$a^{(0)}_{\omega/\rho0 f}
\equiv \frac{A^{(0)}_{\omega f}}{A^{(0)}_{\rho f}}\,.$$
Now, neglecting the difference of phase spaces in the decays,
we can easily describe the measurable quantity
\begin{equation}
r_{\omega/\rho0 f}\equiv \frac{\Gamma(\omega\to f)}
{\Gamma(\rho^0\to f)}=|a_{\omega/\rho0 f}|^2\,.
\end{equation}
Each pair of decays has its own parameter $a^{(0)}_{\omega/\rho0 f}$,
while  $\wtG$ is universal.

For decays $(\omega,\rho^0)\to \pi^+\pi^-$ we can assume absence of
direct isospin violation, {\it i.e.}, $a^{(0)}_{\omega/\rho0(2\pi)}=
0$, and obtain the simple relation
\begin{equation}
r_{\omega/\rho0(2\pi)}\equiv \frac{\Gamma(\omega\to2\pi)}
{\Gamma_\rho}= \frac14\,|\wtG|^2\,,
\end{equation}
where deviation of $K$ from unity
has been neglected. On the basis of Tables~\cite{PDT2} it gives
\begin{equation}
|\wtG|=(6.2\pm0.5)\cdot10^{-2}\,.
\end{equation}
Large variation of this quantity when extracting it from the
consequent issues of the Particle Data Tables~\cite{PDT2,PDT0}
shows that the realistic error should be taken at least twice higher.

Photonic decays $(\omega,\rho^0)\to \pi^0\gamma,\,\eta\gamma,\,e^+e^-$
and $\eta'\to(\omega,\rho^0)\gamma$ contain the real or virtual photon
and have, therefore, nonvanishing direct isospin violation. It can be
assumed, however, to have very simple form based on the structure of
the photon coupling to light quarks:
$$e_u\,\ovl uu\,+e_d\,\ovl dd\,=\frac{e_u+e_d}{\sqrt2}\,
\frac{\ovl uu+\ovl dd}{\sqrt2}+\frac{e_u-e_d}{\sqrt2}\,
\frac{\ovl uu-\ovl dd}{\sqrt2}\,.$$
Evidently, the effective isovector charge is 3 times more than
the isoscalar one. We can append this fact by assumption that
$\ovl uu$ and $\ovl dd$ components of the mesons produce the same
matrix elements. Then for the ratios
$$r_{\eta'\rho0/\omega}\equiv\frac{\Gamma(\eta'\to\rho^0\gamma)}
{\Gamma(\eta'\to\omega\gamma)}\,,~~~
r_{\rho0/\omega\eta}\equiv\frac{\Gamma(\rho^0\to\eta\gamma)}
{\Gamma(\omega\to\eta\gamma)}\,,~~~
r_{\rho0/\omega(ee)}\equiv\frac{\Gamma(\rho^0\to e^+e^-)}
{\Gamma(\omega\to e^+e^-)}$$
we obtain the same expression
\begin{equation}
r=9\left|\frac{1-\frac16\,\wtG}{1+\frac32\,\wtG}\,\right|^2\,.
\end{equation}
A given value of $r$ corresponds to a circle in the complex plane of
$\wtG$, which should intersect another circle, related to eq.(12), and
determine $\wtG$ up to the sign of Im$\,\wtG$.

Data of Tables~\cite{PDT2} provide the values
\begin{equation}
r_{\eta'\rho0/\omega}=9.74\pm1.05\,,~~~
r_{\rho0/\omega\eta}=10.3\pm2.6\,,~~~
r_{\rho0/\omega(ee)}=11.42\pm0.42\,,
\end{equation}
which do not contradict each other. Experimental errors transform
all the corresponding circles into circular bands shown in fig.1.
Though the errors are large, the picture looks consistent with
the value of  $\wtG$ being universal for various decays and
having Re$\,\wtG<0$.

In the same approach we can write
\begin{equation}
r_{\rho0/\rho\pm\pi}\equiv\frac{\Gamma(\rho^0\to\pi^0\gamma)}
{\Gamma(\rho^{\pm}\to\pi^{\pm}\gamma)}
=\left|1-\frac32\,\wtG\,\right|^2\,,
\end{equation}
which shows that interference of direct and cascade contributions
may either suppress or enhance the neutral radiative decay.
Re$\,\wtG<0$ leads to enhancement of the neutral {\it vs.} charged
decay, in agreement with experimental evidences~\cite{PDT2}. This
demonstrates both the role of mixing in pairs of $(\rho^0,
\omega)$-decays and consistency of the dicussed approach to
description of the isospin violation.

The considered photonic decays can be easily described in the framework
of the additive quark model. Its simplest form provides exactly the
same expression as in eq.(14). To check that they have more general
meaning we can consider many-particle decays $(\rho^0,\omega)\to
\pi^0\pi^0\gamma$ which are not easy for application of the
additive quark model. However, we can use the fact of "isotopic
separation" in these decays: only isovector (isoscalar) component
of the photon would participate in the decay of $\rho^0$ ($\omega$)
in the absence of some additional isospin violation, because of mixing 
or any direct effects. Therefore, the quantity 
\begin{equation} 
r_{\rho0/\omega(\pi\pi)}\equiv\frac{\Gamma(\rho^0\to
\pi^0\pi^0\gamma)}{\Gamma(\omega\to\pi^0\pi^0\gamma)}
\end{equation}
should satisfy the same eq.(14). Experimentally~\cite{PDT2},
$r_{\rho0/\omega(\pi\pi)}\approx11\,.$ Uncertainty is still large,
but we see just the expected tendency ($r_{\rho0/\omega(\pi\pi)}$
seems to be higher than the unmixed numerical value of 9).

Up to now we have neglected any really direct violation of
the isospin symmetry. In photonic decays this meant that
the arising matrix elements were assumed to be the same for
$u$- and $d$-components of the mesons, and violation emerged
only due to difference of $e_u$ and $e_d$. However, the slight
difference of $r_{\eta'\rho0/\omega}$ and $r_{\rho0/\omega(ee)}$
may be viewed as an evidence for existence of some additional
direct violation, giving different matrix elements for $u$- and
$d$-components. Another possible evidence for such violation comes
from the ratio
\begin{equation} r_{\omega/\rho\pm\pi}\equiv
\frac{\Gamma(\omega\to\pi^0\gamma)}
{\Gamma(\rho^\pm\to\pi^\pm\gamma)}\,,
\end{equation}
which experimentally~\cite{PDT2} equals $(10.9\pm1.3)$. This
exceeds expectation based on the expression
\begin{equation}
r_{\omega/\rho\pm\pi}
=9\left|1+\frac16\,\wtG\,\right|^2\,,
\end{equation}
with $\wtG\,$ satisfying eq.(13) and having Re$\,\wtG<0\,$.
Sources of additional (direct) violations are still
to be discussed.

\section {Conclusion}

The above examples demonstrate that the $(\rho,\,\omega)$-mixing
reveals itself not only in decays $\omega\to\pi^+\pi^-$ and
$\rho^0\to\pi^0\gamma\,.$ It also affects all pairs of
decays of $\rho^0,\,\omega$ to the same final state and decays
of heavier particles with production of $\rho^0,\, \omega\,$.

Though the current precision is still insufficient for firm
conclusions, existing data on radiative decays of
$(\rho^0,\,\omega)$ and decays to $e^+e^-$ are shown to agree
with the regular, correlated manner expected for the influence
of mixing. The whole set of decays gives additional indirect
support for enhancement of $\rho^0\to\pi^0\gamma$ in comparison
with $\rho^{\pm}\to\pi^{\pm}\gamma\,$.

Future, more precise measurements of those and other decays
will help to separate isospin violation due to $(\rho,\,
\omega)$-mixing from direct violation in various processes
and to study them in detail. This will allow to pick out the
underlying physics and construct adequate models for isospin
violation.

\section* {Acknowledgement}

The work was supported in part by the RFBR grant 00-15-96610.


\newpage

\begin{figure}
\centerline{\epsfig{file=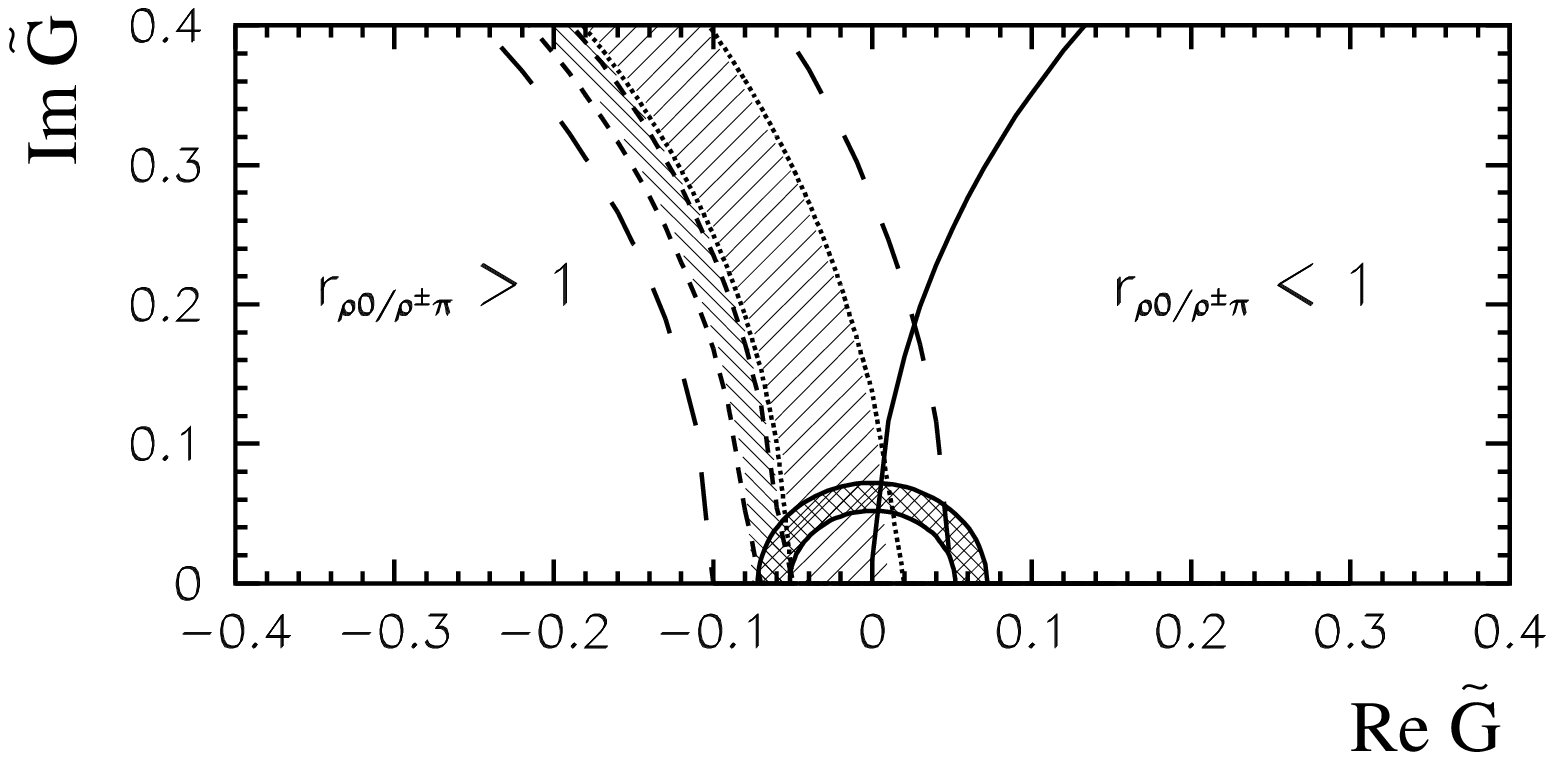,width=16cm}}
\caption{Properties of various $(\rho^0,\omega)$ decay
pairs as seen on the complex plane of $\wtG$ when using
values (15). The long-dashed uncovered band is for
$(\rho^0,\omega)\to\eta\gamma$; the short-dashed band with
left-inclined hatching is for $(\rho^0,\omega)\to e^+e^-$;
the dotted band with right-inclined hatching is for
$\eta'\to(\rho^0,\omega)\gamma$. The solid ring with double
hatching is for $(\omega,\rho)\to\pi\pi$, eq.(13) with the
doubled error. The area to the left/right of the solid line
corresponds to $r_{\rho0/\rho\pm\pi}$ more/less than unity,
{\it i.e.}, to enhancement/suppression of $\rho^0\to\pi^0\gamma$
in respect to $\rho^{\pm}\to\pi^{\pm}\gamma$.}
\end{figure}

\end{document}